\documentclass[a4paper,11pt]{article}
\usepackage[english]{babel}
\usepackage[T1]{fontenc}
\usepackage{textcomp}
\usepackage{epsfig}
\usepackage{latexsym}
\usepackage{graphicx}
\usepackage{amsfonts,amssymb,bm}
\usepackage{color}
\newtheorem{proposition}{Proposition}

\newcommand{\D}{{\rm d}}

\newcommand{\he}{\hat{e}}

\begin{document}
\title{An extension of Poincaré group based on generalized Fermi-Walker coordinates}
\author{Josep Llosa\\
\small Departament de Física Quàntica i Astrofísica, Institut de Ciències del Cosmos (ICCUB),\\
Universitat de Barcelona, Martí Franquès 1, E08028 Barcelona, Spain}

\maketitle

\begin{abstract}

\noindent
The class of accelerated and rotating reference frames has been studied on the basis of generalized Fermi-Walker coordinates. 
We obtain the infinitesimal transformations connecting any two of these frames and also their commutation relations. 
We thus have an  infinite dimensional extension of the Poincaré algebra and, although it turns out to be Abelian extension, and hence trivial, it is noteworthy that, contrarily to Lorentz boosts, acceleration and rotational boost generators commute with each other and with the generators of Poincaré group as well.\\[2ex]
\end{abstract}

\section{Introduction\label{S0}}
The laws of Newtonian mechanics hold in all inertial reference frames, which are in uniform rectilinear motion with respect to each other. This is known as the  principle of relativity of Galilei. However, this principle of relativity can be extended to arbitrary rigid frames, which are in arbitrary translational and rotational motion with respect to each other, i. e. they are mutually related by coordinate transformations like
$$  x^{\prime i} = R^i_{\,j}(t)  x^j + s^i(t)\,, \qquad \qquad t^\prime = t + t_0 \,,  $$
where $\, R^i_{\,j}(t)$ is an orthogonal matrix and $s^i(t)$ arbitrary functions of time. The laws of Newtonian mechanics have the same form in any of these coordinate systems, provided that the necessary inertial force fields ---dragging, Coriolis, centrifugal, \ldots--- are included.

According to the special theory of relativity the laws of physics hold in all Lorentzian reference frames, the relative motion of any couple of these frames is rectilinear and uniform, and the coordinates in any pair of these frames are connected by a Poincaré transformation. 
Endeavouring to set up a theory of gravity consistent with his theory of relativity, Einstein initially aimed to generalize the theory of relativity to accelerated motions \cite{Einstein1907}, but he soon abandoned this idea in favour of the principle of general covariance. Its invariance group, namely spacetime diffeomorphisms, is  much wider than Poincar\'e group but, as soon Kretschmann pointed out \cite{Kretschmann1917}, \cite{Fock} ``since any theory, whatever its physical content, can be rewritten in a generally covariant form, the group of general coordinate transformations is physically irrelevant'' \cite{Antoci2009}. 
Moreover, in Kretschmann's view, special relativity is the one with the relativity postulate of largest content; indeed, its isometry group is a ten-parameter group, which is the largest isometry group in four dimensions, whereas for generic spacetimes in general relativity the isometry group reduces to the identity.

More recently other authors have insisted in the convenience of restricting general covariance \cite{Ellis1995} and even a Principle of restricted covariance has been explicitly stated \cite{Zalaletdinov1996}. 
There is also in the literature a renewed interest in accelerated reference frames. In a non-relativistic context we should mention the  extensions of Galilei algebra to encompass acceleration \cite{Duval1993}, \cite{Lukierski2007}, \cite{Gomis2012} and, from a relativistic standpoint, the approach by Mashhoon \cite{Mashhoon} in view of its possible application to non-local gravitational theories \cite{Mashhoon2}.

Our aim is to find an extension of the principle of relativity, meaning that there is a class of coordinate systems  ---larger that the Lorentzian class but more restrictive than general curvilinear coordinates-- such that the laws of physics have the same form, including inertial force fields if necessary; in as much the same way as the extended principle of relativity for rigid reference frames holds in Newtonian mechanics, as commented at the start. 

Fermi-Walker (FW) coordinates \cite{MTW}, \cite{Synge1} are characterized by an arbitrary origin worldline, whose proper velocity stands for the time axis, whereas the triad of space axis evolve without rotating in a specific way, namely Fermi-Walker transport \cite{Walker}, just to stay orthogonal to proper velocity. Often these coordinates have been seen as the relativistic generalization of the coordinates associated to an accelerated, non-rotating observer \cite{MTW}, \cite{Bini2015}, \cite{Marzlin1993}.

This class of observers and associated coordinates can be extended by allowing the space axes to have an arbitrary rotational motion \cite{MTW}, \cite{Ni1978}, \cite{Gourgoulon}. We shall refer to these as {\em generalized Fermi-Walker coordinates} (GFW)\footnote{Gourgoulhon \cite{Gourgoulon} uses the term ``coordinates with respect to a generic observer''}.

In section 2 we outline the main features of FW and GFW coordinate systems in Minkowski spacetime and see the specific form of Minkowski interval when written in these coordinates. Then we prove that the latter specific form is exclusive of GFW coordinates and state a kind of uniqueness result associated to them. 

The Minkowski metric components in these coordinates are ten well defined functions, namely $\,g_{\mu\nu}(X^j,T;f_J(T))\,$ that involve six functions of time $\,f_J(T)\,$, $ J = 1 \dots 6$. These functions are connected with the origin proper acceleration and the angular velocity of the spatial triad of axes. 

In passing from one GFW system of coordinates to another the $\,g_{\mu\nu}(X^j,T;f_J)\,$ are form invariant, i.e. considered as functions of the ten variables $X^j$, $T$ and $f_J$ they do not change. However they are not isometries in the proper sense because, if we consider the components as functions of the four coordinates, namely $\,\overline{g}_{\mu\nu}(X^j,T) := g_{\mu\nu}(X^j,T;f_J(T))\,$, their values do change because the six functions $\,f_J(T)\,$ differ from one GFW system to the other. This dual feature incline us to use the term {\em generalized isometries} \cite{Bel93}. 

We then study the infinitesimal transformations connecting two GFW coordinate systems, as the solutions of a generalized Killing equation, and derive the infinitesimal generators . As these transformations imply not only a change in the spacetime coordinates $(X^j, T)$ but also in the six functions $f_J(T)$ characterizing the GFW system, the infinitesimal generators act on a manifold that is much larger than merely Minkowski spacetime (an infinite dimensional one, actually).
These generators span an infinite dimensional extension of Poincaré algebra which includes acceleration and rotation, which may be taken as the mathematical embodying of an extension of the principle of special relativity abiding arbitrary translational and rotational motions.
  
\section{Generalized Fermi-Walker coordinates  \label{S1}}
Let  $\,z^\mu(\tau) \,$ be a timelike worldline in ordinary Minkowski spacetime,  and $u^\mu = \dot{z}^\mu(\tau)\,$ and and $a^\mu = \ddot{z}(\tau) \,$ the proper velocity and acceleration 4-vectors. (We take $c=1$, Greek indices run from 1 to 4 and Latin indices from 1 to 3; $\,x^\mu\,$ refer to Lorentzian coordinates and the summation convention is always understood unless the contrary is explicitely said.)

A 4-vector $\,w^\mu(\tau)\,$ is Fermi-Walker transported \cite{Synge1} along $z^\mu(\tau)\,$ if 
\begin{equation}  \label{E1}
\frac{\D w^\mu}{ \D \tau} = \left(u^\mu a_\nu - u_\nu a^\mu \right)\, w^\nu
\end{equation}

Let us now consider an orthonormal tetrad, $\he^\mu_{(\alpha)}(\tau) \,$, which is FW transported along $z^\mu(\tau) \,$ and such that $\he^\mu_{(4)} = u^\mu \,$. For a given point in spacetime, $\,x^\mu\,$, the Fermi-Walker coordinates \cite{Synge1}, \cite{Gourgoulon} with space origin on $z^\mu(\tau)$ are:
\begin{description}
\item[The time] $T(x^\nu)$,  given as an implicit function by
\begin{equation}  \label{E3}
  \left[x^\mu - z^\mu(T)\right]\,u_\mu(T) = 0
\end{equation}
\item[The space coordinates] $X^i$, defined by
\begin{equation}  \label{E4}
 \hat{X}^i = \left[x_\mu - z_\mu(T(x))\right]\,\he^\mu_{(i)}(T(x))
\end{equation}
\end{description}

FW coordinates are the local coordinates of a non-rotating accelerated observer \cite{MTW}. A natural generalization, that also includes arbitrary rotational motion, is based on the notion of generalized Fermi-Walker (GFW) transport \cite{MTW}, \cite{Ni1978} of a vector $w^\mu$ along the worldline $z^\mu(\tau)$: 
\begin{equation}  \label{E5}
\frac{\D w^\mu}{ \D \tau} = \Omega^\mu_{\;\nu}(\tau) \, w^\nu 
\end{equation}
where 
\begin{equation}  \label{E5a}
 \Omega^\mu_{\;\nu} = u^\mu a_\nu - a^\mu u_\nu - \epsilon^\mu_{\;\nu\alpha\beta}\omega^\alpha u^\beta 
\end{equation}
is the {\em (spacetime) angular velocity} and $\omega^\alpha(\tau)$ is an arbitrary vector that is orthogonal to $u^\mu$ that we shall call {\em proper angular velocity} 4-vector.

Consider now a new tetrad $\,\{{e}^\nu_{\;(\alpha)}(\tau)\}_{\alpha =1\ldots 4}\,$, with $\,{e}^\nu_{\;(4)}(\tau) = u^\nu(\tau)\,$, that is GFW transported along $z^\mu(\tau)$. We shall write:
$$ {e}_{\nu(\alpha)}  =\eta_{\nu\mu} {e}^\nu_{\;(\alpha)} \,, \qquad \qquad {e}^{\nu(\alpha)} = {e}^\nu_{\;(\beta)} \eta^{\beta\alpha}  $$
and the components of the angular velocity in this comoving base are:
\begin{equation}  \label{E5b}
 \hat\Omega^\alpha_{\;\beta} = \Omega^\mu_{\;\nu} e_\mu^{\; (\alpha)} e^\nu_{\; (\beta)} \,, \qquad \quad
 \hat\Omega^4_{\; i} = \hat\Omega^i_{\; 4}  = \hat{a}^i \,, \qquad   \hat\Omega^i_{\; j} = \epsilon^i_{\; jk} \hat\omega^k
\end{equation} 
where $\,\hat\omega^l = \omega_\nu e^\nu_{\;(l)}$\,.

On the basis of the origin worldline $z^\mu(\tau)$ and the space axes ${e}^\mu_{\;(i)}$ we can introduce the generalized Fermi-Walker coordinates of a point $x^\nu$: (a) the time $T(x)$ is defined as in equation (\ref{E3}) and (b) the space coordinates are
\begin{equation}  \label{E6}
 X^j = \left[ x_\mu - z_\mu(T(x))\right]\, {e}^\mu_{\;(i)}(T(x))
\end{equation}
The inverse coordinate transformation --- from GFW to Lorentzian coordinates--- is 
\begin{equation}  \label{E7a}
\left(T,\, X^l\right) \longrightarrow x^\mu \,, \qquad \qquad 
x^\mu = z^\mu(T) + X^l {e}^\mu_{\;(l)}(T) 
\end{equation}
whence it easily follows that
\begin{equation}  \label{E7}
 \D x^\mu = \left[ \left( 1 + \vec{X}\cdot{\vec{a}}\right) \, u^\mu + \epsilon^l_{\;ik} \hat\omega^k X^i {e}^\mu_{\;(l)} \right] \,\D T+ {e}^\mu_{\;(l)}\,\D X^l
\end{equation}
and the Minkowski metric in GFW coordinates is
\begin{equation}  \label{E8}
\D s^2 = \D \vec{X}^2 + 2\, \D T \,\D \vec{X} \cdot \left(\vec{X}\times \vec{\omega} \right) - \D T^2 \,\left[  \left( 1 + \vec{X}\cdot{\vec{a}}\right)^2 - \left(\vec{X}\times \vec{\omega} \right)^2 \right] 
\end{equation}
where the usual standard 3-vector notation has been introduced for the sake of brevity, with $\vec{\omega}=(\hat{\omega}^1,\hat{\omega}^2,\hat{\omega}^3)$ and $ {\vec{a}}= (\hat{a}^1,\hat{a}^2,\hat{a}^3)$.

Had we to compare these GFW coordinates with the ordinary FW coordinates based on the same world line, we should obtain
$ X^j = {R}^j_{\; i}(T) \hat{X}^i \,$, where $ {R}^j_{\; i}(\tau)$ is a rotation matrix satisfying
$$ \dot{R}^j_{\; i}(\tau)  = \epsilon^l_{\;ik} \hat\omega^k(\tau) {R}^j_{\;\, l}(\tau) $$
($\epsilon^l_{\;jk} = \epsilon_{ljk}\,$ is the three dimensional Levi-Civita symbol, regardless the position of the Latin indices).
 
It is worth to remark that, whereas $a^\mu$, $\omega^\nu$ and $\Omega_{\mu\nu}$ are the components of respectively proper acceleration, proper angular velocity vector and spacetime angular velocity on an external Lorentzian coordinate base, 
$\hat{a}^j(\tau)$, $\hat{\omega}^j(\tau)$ and $\hat{\Omega}_{\alpha\beta}(\tau)$ are the components of these objects with respect to the GFW transported base. This is why we shall refer to them as {\em intrinsic} proper acceleration and so on. 

\subsection{The GFW reference frame with origin $z^\mu(\tau)$ and angular velocity $\Omega^\mu_{\;\;\nu}(\tau)$ \label{S1.1} }
Any GFW transported tetrad with angular velocity $\Omega^\mu_{\;\;\nu}(\tau)$ is a solution of the linear ordinary differential system (\ref{E5}-\ref{E5a}). Its general solution is 
\begin{equation}  \label{E9}
{e}^\mu_{(\alpha)}(\tau) = \Lambda^\mu_{\;\;\nu}(\tau)\, {e}^\nu_{(\alpha)}(0)
\end{equation}
where $\,\Lambda^\mu_{\;\;\nu}(\tau)\,$ is a solution of the differential system 
\begin{equation}  \label{E9a}
\dot \Lambda^\mu_{\;\;\nu} = \Omega^\mu_{\;\rho}(\tau) \, \Lambda^\rho_{\;\;\nu} \,, \qquad {\rm with} \qquad 
\Lambda^\mu_{\;\;\nu}(0) = \delta^\mu_\nu 
\end{equation}

Therefore two tetrads, $\, {e}^\mu_{(\alpha)} =  {e}^{\prime \mu}_{(\alpha)}= u^\mu\,$, that are GFW transported along the same worldline with the same angular velocity will only differ in their initial values and as, besides $\, {e}^\mu_{(4)} =  {e}^{\prime \mu}_{(4)}= u^\mu\,$, these initial values are connected by a space rotation 
$$  {e}^{\prime \mu}_{(\alpha)}(0) = \sum_{\beta=1}^4  {e}^\mu_{(\beta)}(0)\,R^\beta_{\;\;\alpha} \,, \qquad {\rm with} \qquad 
R^\beta_{\;\;4} = R^4_{\;\;\beta} =\delta^\beta_4  \,, \qquad $$
$\,\left(R^i_{\;\;j}\right)_{i,j=1\ldots 3}$ being a constant orthogonal matrix. Combining the latter with (\ref{E9}) we have that
\begin{equation}  \label{E10}
 {e}^{\prime \mu}_{(\alpha)}(\tau) =  {e}^\mu_{(\beta)}(\tau)\,R^\beta_{\;\;\alpha} 
\end{equation}

Hence all GFW transported tetrads along a given worldline with the same angular velocity are the same apart from an initial space rotation and, according to the definition  the GFW coordinates based on any of these tetrads will differ at most in a constant rotation:
$$ \tau^\prime = \tau \,, \qquad \qquad X^i =  R^i_{\; j} X^{\prime j}  $$

Given a GFW coordinate system with origin $\,z^\mu(\tau)\,$ and angular velocity $\Omega^\mu_{\;\alpha}(\tau)$, the 3-parameter congruence of worldlines $X^i =$constant, $\tau \in \mathbb{R}$ are the ``history'' of a place in the reference space associated to the GFW coordinates and, as commented above, in any other GFW coordinated system based on   $\,z^\mu(\tau)\,$ and $\Omega^\mu_{\;\alpha}(\tau)$ we shall still have that
$X^{\prime i} = \,\left(R^{-1}\right)^i_{\;j} X^j = {\rm constant} \,$.

The equation defining this 3-parameter congruence is
\begin{equation}  \label{E6a}
\varphi^\mu(T,\, \vec{X}) = z^\mu(T) + X^i\, {e}^\mu_{(i)}(T) \,, \qquad \qquad \tau \in\mathbb{R}  
\end{equation}
and the proper time rate at the place $\vec{X}$ is
$$  \D \tau =  \D T \, \sqrt{\left[1 + \vec{X}\cdot{\vec{a}}(\tau) \right]^2 - \left[\vec{X}\times \vec\omega\right]^2} \,;$$
This is the time ticked by a stationary atomic clock at $\vec{X}$ and it coincides with $T$ at the origin. 
In general, $\tau \neq T$ and usually the readings of proper time $ \tau$ by two stationary clocks at two different places will not keep synchronized. It is thus more convenient to use the {\em synchronous time} $\,T\,$ instead of local proper time.

The proper velocity vector of the worldline $\vec{X} =$constant at the synchronous time $T$ is
\begin{equation}  \label{E10a}
 U^\mu(T,\vec{X}) = \frac{ \left(1 + \vec{X}\cdot{\vec{a}} \right)\,u^\mu + \epsilon^l_{\;\;ik} X^i \hat\omega^k \, {e}_{(l)}^\mu}{\sqrt{\left[1 + \vec{X}\cdot{\vec{a}} \right]^2 - \left[\vec{X}\times \vec\omega\right]^2} } 
\end{equation}
The first term corresponds to the origin translational velocity, whereas the second term reflects the rotational motion.

Due to the presence of a square root in the denominator, the domain of validity of the GFW coordinates is restricted to the region $\, \left|1 + \vec{X}\cdot{\vec{a}} \right| > \left|\vec{X}\times \vec\omega\right| $ and the equality defines the horizon of the GFW coordinate system.

\section{Uniqueness \label{S1.2}}
We now prove that the expression (\ref{E8}) of the metric is exclusive for GFW coordinates in Minkowski spacetime.

\begin{proposition}  \label{p1}
If in some coordinate system $\left( X^i, T\right)$ the spacetime metric has the form (\ref{E8}), then there is a worldline $z^\mu(\tau)$
and an orthonormal tetrad $\,e^\mu_{\;(\alpha)}(\tau)\,$, $\alpha = 1 \ldots 4\,$, such that $\left( X^i, T\right)$ are the GFW coordinates based on that worldline and tetrad in a locally Minkowskian spacetime.
\end{proposition}
\paragraph{Proof:} It is straightforward to check that the Riemann tensor for the metric (\ref{E8}) vanishes; therefore the spacetime is locally Minkowskian.

Then consider the matrix $\hat\Omega^\alpha_{\;\beta}(\tau)$ defined by
\begin{equation}  \label{E21c}
\hat\Omega^i_{\;j}(\tau) = \epsilon^i_{\;j k} \hat\omega^k(\tau) \,, \qquad \qquad 
\hat\Omega^4_{\;i}(\tau) = \hat\Omega^i_{\;4}(\tau) = \hat{a}^i(\tau)  
\end{equation}
where the functions $\hat{a}^i$ and $\hat\omega^k$ are obtained from the coefficients in the metric (\ref{E8}). Then take $e^\mu_{\;(\alpha)}(\tau)$ as the solution of the ordinary differential system
\begin{equation}  \label{E21}
\frac{\D e^\mu_{\;(\alpha)}}{\D \tau} = e^\mu_{\;(\beta)} \hat\Omega^\beta_{\; \alpha}(\tau)
\end{equation}
for some initial data $\{e^\mu_{\;(\alpha)}(0)\}_{\alpha = 1 \ldots 4}$, that form an orthonormal tetrad, 
with $e^\mu_{\;(4)}(0) $ timelike. Due to the particular form of the matrix $\hat\Omega^\beta_{\; \alpha}$, it is obvious that  $e^\mu_{\;(\alpha)}(\tau)$ is an orthonormal tetrad for all $\tau$ as well.

Being $e^\mu_{\;(4)} $ a timelike vector, consider a worldline $z^\mu(\tau)$ such that $\dot{z}^\mu(\tau) = e^\mu_{\;(4)}(\tau) $ and the matrix
$$ \Omega^\mu_{\; \nu} = \hat\Omega^\beta_{\; \alpha} e^\mu_{\;(\beta)} e_\nu^{\;(\alpha)} $$
(indices are raised and lowered with $\eta_{\alpha\beta} = {\rm diag}[+1,+1,+1,-1]\,$).
It is straightforward to see that the tetrad $e^\mu_{\;(\alpha)}(\tau)$ is GFW transported along $z^\mu(\tau)$ with an angular velocity 
$\Omega^\mu_{\; \nu} (\tau) \,$.

Consider finally the coordinate transformation 
$$ \left( X^i, T\right) \longrightarrow x^\mu = z^\mu(T) + X^j\,e^\mu_{\;(j)}(T) $$
whose Jacobian and inverse Jacobian are  respectively given by 
\begin{equation}  \label{E22}
\D x^\mu = \left(\dot{z}^\mu(T) +  e^\mu_{\;(\alpha)}(T)\,\hat\Omega^\alpha_{\;j}(T)\,X^j \right) \,\D T + e^\mu_{\;(j)}(T) \,\D X^j
\end{equation}
\begin{equation}  \label{E23}
\D T = - \frac{e_{\mu(4)} (T)\,\D x^\mu}{1+\vec{X}\cdot\vec{a}(T)} \,, \qquad \qquad 
\D X^i = \left( e_{\mu(i)} + \frac{e_{\mu(4)}(T)\,\hat\Omega^i_{\;j}(T) \,X^j}{1+\vec{X}\cdot\vec{a}(T)}  \right)\,\D x^\mu
\end{equation}
Substituting the latter in the expression (\ref{E8}) we obtain that, in the coordinates $x^\mu$, the invariant interval is $\D s^2 = \eta_{\mu\nu} \,\D x^\mu\,\D x^\nu \,$; hence $x^\mu$ are Lorentzian coordinates and $ X^i,\,T\,$ are the GFW coordinates for the worldline $z^\mu(\tau)$ and the tetrad $e^\mu_{\;(\alpha)}\,$. \hfill $\Box$

\subsection{A geometric characterization of GFW coordinate systems  \label{S1.3}}
We shall see now that, if the hypersurfaces $T = \,$constant are hyperplanes in Minkowski spacetime and $X^j$ are Cartesian coordinates on these hyperplanes, then $(X^j,T)$ are GFW coordinates apart from a shift in the origin.

\begin{proposition}  \label{p2}
Let $(X^j,T)$ be a system of coordinates such that the Minkowski metric spatial components are $g_{ij} = \delta_{ij}$, then: {\bf (a)} the hypersurfaces $T = $constant are flat, {\bf (b)} the rank of their extrinsic curvature $K_{ij}$ is at most one, and {\bf (c)} it satisfies that $\partial_{[i}K_{j]l} = 0 $\,.
\end{proposition}

\paragraph{Proof:} The metric restricted to the hypersurfaces  $T = \,$constant, $\overline{g}_{ij} = \delta_{ij}$, is flat and the ambient metric is Minkowski metric. Then, if $K_{ij}$ is the second fundamental form, Gauss equation \cite{Hicks} implies that $K_{i[j}K_{l]k}=0\,$.

The latter equation having the same symmetries as a Riemann tensor and being three the number of effective dimensions, it is equivalent to its trace $(il)$, that is:
$$ K^j_{\;i} K^i_{\; k} -  K^i_{\;i} K^j_{\; k} = 0 $$
which implies that  $K^j_{\;i}$ has two eigenvalues, namely $\,K^i_{\;i}\,$ (simple) and 0 (double), hence ${\rm rank}\,K^j_{\;i} \leq 1$.

The relation $\partial_{[i}K_{j]l} = 0 \,$ is a consequence of the Codazzi-Mainardi equation \cite{Hicks} and the fact that $X^j$ are Cartesian coordinates for the first fundamental form on the hypersurfaces  $T = $constant. \hfill $\Box$

\begin{proposition}   \label{p3}
If the metric spatial components are $g_{ij} = \delta_{ij}$ and the second fundamental form on $T = $constant vanishes, then $(X^j,T)$ are GFW coordinates apart from a shift in the origin.
\end{proposition}

\paragraph{Proof:} Let us define $ v_i = g_{4i} $ and  $N^2 = \vec{v}^{\,2} - g_{44}$, then 
the inverse spacetime metric components are:
$$ g^{ij} = \delta_{ij} - N^{-2} v_ i v_j \,, \qquad g^{4i} = N^{-2} v_i \,, \qquad g^{44} = - N^{-2} = \det(g^{\mu\nu}) \neq 0$$
and the connexion symbols are:
\begin{equation}  \label{E231}
\left. \begin{array}{lll}
 \{ij | k\} = 0 \,, \qquad & \{ ij| 4\} = \partial_{(i} v_{j) \,,} \,, \qquad & \{4 i| j\} =\partial_{[i} v_{j]} \\[2ex]
\multicolumn{2}{l}{ \{44 | i\} = \partial_T v_i + N\,\partial_i N + 2 \,\vec{v}\cdot \partial_i \vec{v} } \,, \quad & 
  \{4 4| 4\} = N\,\partial_T N + 2 \,\vec{v}\cdot \partial_T \vec{v} 
   \end{array}    \right\} 
\end{equation}
The unit covector normal to the hypersurfaces $T = \,$constant  and the second fundamental form are respectively
$$ n_a = - N\,\delta^4_a \qquad {\rm and} \qquad 
 K_{ij} = - N \,\Gamma_{ij}^4 = N^{-1} \,\partial_{(i} v_{j)} $$
The vanishing of $K_{ij}$ then implies that $\,\partial_i v_j = W_{ij}$ is skewsymmetric, whose integrability conditions,
$\partial_l W_{ij} = \partial_i W_{lj}$, combined with by the Jacobi identity, imply that $\partial_j W_{il} = 0\,$, that is 
\begin{equation}  \label{E232}
W_{ij} = \epsilon_{ijk} \hat\omega^k(T)\qquad \mbox{ and therefore } \qquad 
v_i = \epsilon_{ijk} X^j\hat\omega^k(T) + V_i(T) 
\end{equation} 
So far we have used the restrictions imposed by Gauss and Codazzi-Mainardi equations. We have still to exploit the vanishing of the components  $R_{4i4j}\,$. Including that $\partial_{(i} v_{j)} = 0$ and the connexion symbols (\ref{E231}), we arrive at
$$ R_{4i4j} = 0 \qquad \Leftrightarrow \qquad \partial_{ij} N = 0 \,, $$
that is 
$$ \exists \;\; \vec{a}(T) \quad{\rm and} \quad B(T) \quad\mbox{such that}\qquad N  = B + \vec{X}\cdot\vec{a} $$
and, provided that $B\neq 0\,$, the time coordinate $T$ can be redefined so that $B = 1$.

Therefore, in these coordinates the Minkowski metric reads
\begin{equation}  \label{E233}
\D s^2 = \D \vec{X}^2 + 2\,\D T\, \D \vec{X} \cdot \left( \vec{X}\times \vec\omega + \vec V \right) + \D T^2\,\left(-\left[1 + \vec{X}\cdot\vec{a}\right]^2 + \left[ \vec{X}\times \vec\omega + \vec V \right]^2 \right)  
\end{equation}
Finally, if we shift the origin as: $\,\tilde{X}^j = X^j + M^j(T) \,$, where $\vec{M}(T)$ is a solution of 
$\partial_T{\vec{M}} + \vec\omega\times \vec{M} = \vec{V} \,$, the Minkowski spacetime interval in the new coordinates has the form (\ref{E8}), i. e. $(\tilde{X}^j, T)$ are GFW coordinates. \hfill$\Box$
  
\section{Generalized isometries \label{S2}}
Deriving a closed expression for the transformation relating two different GFW coordinate systems would imply to invert the transformation law (\ref{E6a}), which is not feasible in general. However we can obtain expressions for infinitesimal transformations with the help of the notion of generalized isometry \cite{Bel93}.

In a GFW coordinate system the invariant interval has a very specific shape (\ref{E8}) that involves six arbitrary functions $\hat{a}^i( T)$  and $\hat{\omega}^j( T)$. The transformation formulae relating two GFW coordinate systems, 
$\, X^\mu =(\vec{X}, T) \longrightarrow X^{\prime \mu} =(\vec{X}^\prime , T^\prime)\,$, must preserve this overall shape but, perhaps, with a different sextuple of functions, $\left(\hat{a}^{\prime i}(t), \hat{\omega}^{\prime j}(t)\right)\,$. We shall call this transformation a {\em generalized isometry} \cite{Bel93} because the interval is:
\begin{equation} \label{E11}
 \D s^2 = g_{\alpha\beta}(X^\nu, f_I(X)) \D X^\alpha\,\D X^\beta = g_{\alpha\beta}(X^{\prime \nu}, f^\prime_I(X^\prime)) \D X^{\prime \alpha} \,\D X^{\prime \beta} 
\end{equation}
Although the functions $g_{\alpha\beta}(X,f_I)$ are the same, the metric coefficients, $\overline{g}_{\alpha\beta}(X)=g_{\alpha\beta}(X,f_I(X))$, are different because the values $f_I(X)$ change to $f^\prime_I(X^\prime)$ in passing from one system to the other. This is the reason why the transformation is not actually an isometry and we need to introduce the notion of generalized isometry.

In the present case $f_I$  are six function that only depend on the coordinate $T$ and can be arranged as the skewsymmetric matrix
\begin{equation}   \label{E10b}
\hat\Omega_{\alpha\beta} = 
\left( \begin{array}{c|c}
    \epsilon_{ijk} \hat\omega^k(T) & \hat{a}^i(T) \\
		\hline
		- \hat{a}^j(T) & 0
		\end{array}  \right)  
\end{equation}

Consider now the infinitesimal transformation
\begin{equation} \label{E12}
 X^{\prime \alpha} = X^\alpha + \varepsilon \,\xi^\alpha(X) \,, \qquad f^\prime_I(X) = f_I(X) + \varepsilon \,\Phi_I(X) 
\end{equation}
and therefore $\, f^\prime_I(X^\prime) = f_I(X) + \varepsilon \,\left[\Phi_I(X) + \xi^\alpha\,D_\alpha f_I(X) \right]\,$, where $X^4 = T$ and $D_\alpha = \frac{\partial\;\;}{\partial X^\alpha}\,$.

Substituting this in equation (\ref{E11}) and keeping only first order terms we obtain
\begin{equation} \label{E13a}
   \xi^\alpha\,D_\alpha \overline{g}_{\mu\nu} + 2 D_{(\nu}\xi^\alpha  \overline{g}_{\mu)\alpha} + G_{\mu\nu}(X) = 0 
\end{equation}
where 
$$  \overline{g}_{\mu\nu}(X) = g_{\mu\nu}(X, f_I(X))  \qquad {\rm and}  \qquad 
G_{\mu\nu}(X) = \sum_I \Phi_I \,\left(\frac{\partial g_{\mu\nu} }{\partial f_I}\right)_{(X, f_I(X))} $$

We have as many functions $\Phi_I$ as $f_I$ and they can be also arranged in the skewsymmetric matrix $\hat{F}_{\alpha\beta}$ as
\begin{equation} \label{E10c}
 \hat{F}_{ij} = \epsilon_{ijk} \hat\alpha^k(T) \,, \qquad \qquad \hat{F}_{i4} = - \hat{F}_{4i} = \hat{A}^i(T)  
\end{equation}
in much the same way as we did for the $f_I$ in the matrix (\ref{E10b}).

Equation (\ref{E13a}) can also be written as the {\em generalized Killing equation}
\begin{equation} \label{E13}
\overline \nabla_\mu \xi_\nu + \overline \nabla_\nu \xi_\mu  + G_{\mu\nu}  = 0 \,,
\end{equation}
where $\overline\nabla $ is the Levi-Civita connexion for $\overline{g}_{\mu\nu}$ and $\xi_\mu = \overline{g}_{\mu\nu} \xi^\nu \,$.

Notice that the infinitesimal transformation (\ref{E12}) acts on $X^\alpha$ and also on the functions $\hat\Omega_{\alpha\beta}$; hence the arena to represent their action is not Minkowski spacetime but rather the larger (infinite dimensional) manifold
$$ \mathcal{M} \subset \left\{\left(\vec{X},T, \hat{\Omega}_{\alpha\beta}(t) \right) \in \mathbb{R}^4 \times \mathcal{C}^0\left(\mathbb{R},\mathbb{R}^6\right)\,, \mbox{such that } \left| 1 + \vec{X}\cdot \vec{a}(T)\right| > \left|\vec{X}\times \vec\omega(T)\right|
 \right\} $$ 
(the inequality is to ensure that the metric (\ref{E8}) is non-degenerate). The infinitesimal generator then looks like:
\begin{equation} \label{E14}
\xi^\alpha\,D_\alpha +  \int_\mathbb{R} \D t\, \hat{F}_{\alpha\beta}(t)\, \frac{\delta \;\;}{\delta \hat{\Omega}_{\alpha\beta}(t)} 
\end{equation}
where $\,\xi^\alpha $ depends on $X^j$, $T$ and $\hat{\Omega}_{\mu\nu}(t)$.

The need of the additional information contained in the six functions $\,\hat{\Omega}_{\alpha\beta}(t)\,$ comes from the fact that the coordinates $\,(\vec{X},T)\,$ are not enough to determine an event in spacetime unless we further indicate the family of GFW observers to which these coordinates belong. 
To fix $\hat{\Omega}_{\alpha\beta}(t)$ means choosing a subclass of GFW coordinate systems, that corresponding to GFW observers with the same intrinsic spacetime angular velocity. Each one is characterized by the initial values of the tetrad, $e^\mu_{(\alpha}(0)$, and the initial point in the origin worldline, $z^\mu(0)$, (the tetrad $e^\mu_{(\alpha}(\tau)$ and the worldline $z^\mu(\tau)$ are then obtained as in the proof of Proposition \ref{p1} in section \ref{S1.2}). 
By the way, it is the same kind and number of parameters as for Poincaré group, although they are not Lorentzian observers because $\hat{\Omega}_{\alpha\beta}(t) \neq 0$. 

The 4-dimensional submanifolds $\,\mathcal{M}_{\hat\Omega} =  \left\{\left(\vec{X},T, \hat{\Omega}_{\alpha\beta}(t) \right) \in \mathcal{M}\right\} \,$ for a given $\hat{\Omega}_{\alpha\beta}(t)$ corresponds to the part of Minkowski spacetime which is accessible to the GFW observers of the class determined by this intrinsic spacetime angular velocity. 

From a physical viewpoint it is worth to remark here that the magnitudes $\,\hat{\Omega}_{\alpha\beta}\,$ are measurable by the GFW observer by means of accelerometers ($\,\hat{a}_j= \hat{\Omega}_{4j}\,$) and gyrometers ($\,\hat{\Omega}_{jk}\,$) along the observer's spatial axes. As their measures do not need resorting to anything external, these magnitudes are ``absolute'', hence there is no sense in referring to them as ``relative acceleration'' or ``relative angular velocity''.

From the interval expression (\ref{E8}) we have that 
\begin{equation} \label{E14a}
 g_{ij} =0 \,, \qquad g_{4i} = \epsilon_{ijk} X^j \hat\omega^k \,, \qquad g_{44} = \left(\vec{X}\times\vec\omega\right)^2 -\left(1 +  \vec{X}\cdot \vec{a} \right)^2 
\end{equation}
and, as it follows from (\ref{E13a}), 
\begin{equation} \label{E14b}
  G_{ij} =0 \,, \qquad G_{4i} = \epsilon_{ijk} X^j \hat\alpha^k \,, \qquad \frac12\,G_{44} =  - \left(1 +  \vec{X}\cdot \vec{a} \right) \,\vec{X}\cdot \vec{A} +  \left(\vec{X}\times\vec\omega\right)\cdot\left(\vec{X}\times\vec\alpha\right) 
\end{equation}

Using this and after a little algebra, the generalized Killing equation splits in several blocks as
\begin{eqnarray} \label{E15a}
\mbox{{\bf space:}} & \qquad & D_{(i} \xi_{j)} = 0 \\   \label{E15b}
\mbox{{\bf cross:}}\, & \qquad & \displaystyle{ -\left(1 +  \vec{X}\cdot \vec{a}\right)^2 D_{i} \xi^4 + D_{4} \xi_i + \epsilon^j_{\;lk} X^l \hat\omega^k D_i \xi_j -  \xi_j \epsilon^j_{\;ik} \hat\omega^k +\epsilon_{ilk}X^l \hat\alpha^k = 0 }
		\\[2ex]  
\mbox{{\bf time: }} & \qquad & \displaystyle{-\left(1 +  \vec{X}\cdot \vec{a}\right)\,D_4 \left[ \left(1 +  \vec{X}\cdot \vec{a}\right)\,\xi^4 \right] +  \epsilon^j_{\;lk} X^l \hat\omega^k D_4 \xi_j +}  \nonumber \\[2ex]     \label{E15c}
 & & \qquad \displaystyle{ \frac12\,D_jg_{44} \left(\xi_j - \xi^4 \epsilon_{jlk} X^l \hat\omega^k\right) + \frac12\, G_{44} } = 0 
\end{eqnarray}

The general solution to the space block is 
\begin{equation} \label{E16}
\vec\xi = \vec{f}(T) + \vec{X} \times\vec{g}(T) \,, \qquad {\rm where } \qquad \vec\xi =(\xi_1, \xi_2, \xi_3)   \,,
\end{equation}
which, substituted in the cross equation yields
\begin{equation} \label{E17}
 -\left(1 +  \vec{X}\cdot \vec{a}\right)^2 D_{i} \xi^4 + \dot\phi_i + \epsilon_{ilk} X^l \gamma^k =0 \,, 
 \end{equation}
where a ``dot'' means derivative with respect to $T$, and 
$$ \vec\phi = \dot{\vec{f}} + \vec{f} \times \omega \,, \qquad  \qquad \vec\gamma = \dot{\vec{g}} + \vec{g}\times\vec\omega + \vec\alpha 
$$  

Equation (\ref{E17}) giving all spatial derivatives of $\xi^4$, it carries some integrability conditions which after some algebra reduce to:
\begin{equation} \label{E18}
\dot{\vec{g}} + \vec{g}\times\vec\omega + \vec\alpha - \vec\phi \times \vec{a} = 0
\end{equation}
In such a case, the general solution of equation (\ref{E17}) is
\begin{equation} \label{E19}
\xi^4 = h(T) + \frac{\vec{X} \cdot \vec\phi }{1 + \vec{a} \cdot \vec{X}} 
\end{equation}

If we now substitute equations (\ref{E16}) and (\ref{E19}) in the time block (\ref{E15c}), we obtain
\begin{equation}  \label{E20}
-\left(1 +  \vec{X}\cdot \vec{a}\right)\,\left[\dot{h} + \vec{a}\cdot\vec{f} + 
\vec{X}\cdot \left( \dot{\vec\phi} +\dot{h}\vec{a} + h\dot{\vec{a}} +  h\,\vec{a}\times\vec\omega + \vec\phi\times\vec\omega +
\vec{g}\times\vec{a} + \vec{A}  \right) \right] = 0 
\end{equation}
where (\ref{E14a}) and (\ref{E14b}) have been included. Putting $\vec \psi = \vec\phi + h \vec{a}\,$, this amounts to
\begin{eqnarray}  \label{E21a}
 \dot{h} + \vec{a}\cdot\vec{f}  = 0  & \qquad &  \\ \label{E21b}
 \dot{\vec\psi} + \vec\psi \times\omega + \vec{g}\times\vec{a} + \vec{A} = 0\,,& \quad{\rm with} \quad &  \vec\psi = \dot{\vec{f}} + \vec{f} \times\omega + h\,\vec{a} 
\end{eqnarray}
Together with (\ref{E18}), these equations constitute an ordinary differential system on the functions $h$, $\vec{f}$ and $\vec{g}$ that occur in the expressions (\ref{E16}) and (\ref{E19}) for $\xi_i$ and $\xi^4$. The solution is obtained in Appendix A  in terms of 4-dimensional variables, namely the 4-vector $f^\alpha(T) = \left( \vec f,\,h \right)$ and the skewsymmetric tensor  $M_{\alpha\beta}(T)$ formed with $\vec\psi\,$  and $\,\vec g\,$ as the electric and magnetic parts respectively. The solutions (\ref{A8}) and (\ref{A9}) depend on ten constant parameters,  $f_0^\mu$ and $M^0_{\alpha\beta}\,$, plus six arbitrary one-variable functions, $\,\hat{F}_{\alpha\beta}(t)\,$. Introducing then these solutions in the expressions (\ref{E19}) and (\ref{E20}), we have that the infinitesimal generator is
\begin{equation}  \label{E22a}
 \left[  f^\mu(T) + M^\mu_{\; \; j} (T) X^j  \right]\,\hat{D}_\mu + \int_\mathbb{R} \D t \,\hat{F}_{\alpha\beta}(t)\, \frac{\delta\quad}{\delta \hat\Omega_{\alpha\beta}(t)} 
\end{equation}
where
\begin{equation}  \label{E22b} 
\hat D_i = D_i \qquad {\rm and} \qquad \hat D_4 = \frac{1}{ 1 + \vec{X}\cdot \vec a(T)} \,\left(D_4 -\epsilon^i_{\;jk} X^j \hat\omega^k(T) \,D_i \right)   
\end{equation}
i. e.  $\hat D_\mu $ are a sort of {\em orthonormalized partial derivatives}.

\subsection{Infinitesimal generators and commutation relations}
As the generator depends on ten constant parameters, namely $f_0^\mu$ and $M^0_{\alpha\beta}\,$, and six arbitrary functions $\hat{F}_{\alpha\beta}(t)$, we can separate this dependence as
\begin{equation}  \label{E23aa}
 f_0^\mu\,\mathbf{P}_\mu + \frac12\, M_0^{\alpha\beta}\,\mathbf{J}_{\alpha\beta} +  \frac12\, \int_\mathbb{R} \D t \,\hat{F}_{\alpha\beta}(t)\,  \mathbf{D}^{\alpha\beta}_{(t)} 
\end{equation}
where $\,M_0^{\alpha\beta} = M^0_{\mu\nu} \eta^{\alpha\mu} \eta^{\beta\nu}\,$ and 
\begin{eqnarray}  \label{E23a}
\mathbf{P}_\mu & = & \Lambda_\mu^{\;\;\nu}(T)\,\hat D_\nu  \,, \qquad \qquad 
\mathbf{J}_{\alpha\beta}= - 2\,k_{[\alpha}(T,\vec X)\,\mathbf{P}_{\beta]}  \\[2ex]  \label{E23c}
\mathbf{D}^{\alpha\beta}_{(t)} & = &  \frac{\delta\quad}{\delta \hat\Omega_{\alpha\beta}(t)} - 2 \,\chi(t,T)\, \Lambda^{\mu [\alpha}(t)\,\Delta^{\beta]}\,(T,t,\vec X) \,\mathbf{P}_\mu 
\end{eqnarray}
with $\quad \chi(t,\tau)= \theta(t)\theta(\tau-t) - \theta(-t)\theta(t-\tau)\,$ and  
\begin{eqnarray} \label{E23d}
k_\beta(T,\vec X) & = & \,\Lambda_{\beta j}(T)\,X^j  + \int_0^T \D t^\prime\, \Lambda_{\beta 4}(t^\prime) 
\\[2ex]  \label{E23e}
\Delta^\nu \left(T,t,\vec X\right) &=& k_\beta(T,\vec X)\,\Lambda^{\beta \nu}(t) - \int_0^t \D t^\prime\, G_4^{\;\;\nu}(t^\prime,t)
\end{eqnarray}
(The matrices $\,\Lambda_\mu^{\;\;\nu}(T)\,$ and $\,G^{\;\;\mu}_\nu(T,t)\,$ are defined in Appendix A.)

In order to better understand how these infinitesimal generators act on the manifold $\mathcal{M}$, we should think of it as sliced in the 4-dimensional submanifolds $\,\mathcal{M}_{\hat\Omega}\,$, each one characterized by a definite choice of the six functions $\hat{\Omega}_{\alpha\beta}(t)$. Any of these submanifolds is coordinated by $(\vec{X}, T)$ and endowed with the metric (\ref{E8}) and corresponds to the part of Minkowski spacetime that is accessible to the class of GFW observers defined by the given $\hat{\Omega}_{\alpha\beta}(t)$.
The generators $\mathbf{P}_\mu$ and $\mathbf{J}_{\alpha\beta}$ act on ---are tangent to--- each slice and span the realization of Poincaré algebra for that particular class of GFW observers. On their turn, the generators $\mathbf{D}^{\alpha\beta}_t$ are transversal to the slicing and are connected with changes in the intrinsic proper acceleration and angular velocity of the observer.

Although the derivation of the Lie brackets between pairs of infinitesimal generators is tedious and intricated, it presents no conceptual subtlety and we shall not derive them explicitely here. An outline of their derivation is postponed to Appendix B. The commutation relations are:
\begin{equation}  \label{E24}
\left.
\begin{array}{lll}
\left[\mathbf{P}_\mu,\mathbf{P}_\nu \right] = 0\,, & \qquad \left[\mathbf{J}_{\alpha\beta},\mathbf{P}_\mu  \right] = 2 \,\eta_{\mu[\alpha} \mathbf{P}_{\beta]}  \,, &  \qquad \left[\mathbf{J}_{\alpha\beta},\mathbf{J}_{\mu\nu}  \right] = 2\,\eta_{\mu[\alpha} \mathbf{J}_{\beta]\nu} - 2\,\eta_{\nu[\alpha} \mathbf{J}_{\beta]\mu}   \\[2ex]
\left[\mathbf{D}^{\alpha\beta}_{(t)},\mathbf{P}_\mu  \right] = 0 \,,  & \qquad   \left[\mathbf{D}^{\alpha\beta}_{(t)},\mathbf{J}_{\alpha\beta} \right] = 0   \,,& \qquad  
 \left[\mathbf{D}^{\alpha\beta}_{(t)},\mathbf{D}^{\kappa\lambda}_{(t^\prime)}  \right] =  0
\end{array}   \right\}
\end{equation}
Thus, the algebra of the infinitesimal transformations connecting generalized Fermi-Walker coordinate systems is an abelian extension of Poincaré algebra.

\section{Conclusion}
We have introduced a class of reference frames with arbitrary translational and rotational motion, namely generalized Fermi-Walker frames. Each one is determined by the worldline of its spatial origin and a triad of spatial comoving axes with an arbitrary rotational motion.

Each GFW system of coordinates, $(T, X^1,X^2,X^3)$ is characterized by six functions of proper time, $\hat{a}_i(\tau)$ and $\hat{\omega}_l(\tau)$, respectively the components of the [proper] acceleration of the origin and the angular velocity of the spatial triad with respect to the comoving axes. These quantities, which are better handled as the skewsymmetric matrix $\hat\Omega_{\alpha\beta}(\tau)$ ---see equation (\ref{E10b})--- are measurable from inside the frame, i. e. without referring to anything external, by means of accelerometers and gyrometers. 

The transformations connecting the coordinates of any pair of frames in the GFW class preserve the form (\ref{E6}) of the spacetime interval, maybe with different functions $\hat\Omega_{\alpha\beta}(\tau)$. Thus we refer to these transformations as {\em generalized isometries}. Infinitesimal generalized isometries satisfy the generalized Killing equation (\ref{E17}), whose solution is an infinite dimensional Lie algebra that contains Poincaré algebra and acceleration boosts plus rotational motions as well. A close look at the commutation relations reveals that it is an Abelian extension of Poincaré algebra.
From the mathematical standpoint, the resulting structure, namely the direct sum of Poincaré algebra and an infinite dimensional abelian algebra, is rather trivial and dull. However it is somewhat surprising because, just as velocity boosts $\mathbf{J}_{4j}$ do not commute with each other nor with rotations $\mathbf{J}_{ij}$ or translations $\mathbf{P}_\mu$, one would expect something similar for acceleration boosts and angular velocity generators, i. e. $\mathbf{D}^{\alpha\beta}_t$, but the effective calculation leads to the ``counterintuitive'' relations (\ref{E24}).

Perhaps the clue of this unexpected commutativity lies in the fact that the parametrization $\hat{\Omega}_{\alpha\beta}(t)$ we have chosen for the extension of Poincaré algebra is rather intrinsic. Indeed, $\hat{\Omega}_{4i}(t)$ and $\hat{\Omega}_{jk}(t)$ are, respectively, the components of proper acceleration and angular velocity in the triad of spatial axes carried by the GFW observer, with reference to nothing external. We might have chosen another parametrization, e. g. the skewsymmetric matrix $\Omega_{\mu\nu} $ in a Lorentzian spacetime basis. This would have resulted in a recombination of generators and, perhaps, a set of commutation relations less simple than (\ref{E24}), but this would not mean a different algebra structure.

We must also remark that the notion of generalized isometry \cite{Bel93} permits to go beyond Kretschmann's idea that, since special relativity admits the widest isometry group, it contains the largest relativity postulate. Our approach here has led to an intermediate group, namely the group of generalized isometries of the interval (\ref{E6}), which is larger than Poincaré group but much smaller than the whole diffeomorfism group. On the other hand, this intermediate group can be seen as the special relativistic counterpart of the Galilei group extensions considered elsewhere \cite{Duval1993},\cite{Lukierski2007},\cite{Gomis2012}.

\section*{Acknowledgement}
The present work is supported by Ministerio de Economía y competitividad (Spanish Gov.) project nş FPA2013-44549-P

\section*{Appendix A}
We here solve equations (\ref{E18}), (\ref{E21a}) and (\ref{E21b}), that we first arrange as
\begin{equation}   \label{A1}
\left.  \begin{array}{l}
        \dot{\vec f} + h\,\vec a + \vec{f}\times\vec\omega = \vec\psi \\[2ex]
        \dot h + \vec f\cdot \vec a = 0
        \end{array}   \right\}
\qquad 
\left.  \begin{array}{l}
        \dot{\vec\psi} + \vec\psi\times\vec\omega + \vec{g} \times \vec a = -\vec A \\[2ex]
        \dot{\vec g} - \vec\psi \times \vec a + \vec{g} \times \vec\omega = -\vec\alpha
        \end{array}   \right\}
\end{equation}
If we put $f^\alpha= \left( \vec f,\,h \right) $ and $\psi^\alpha = \left( \vec\psi,\,0 \right) $, the first pair of equations can be written as
\begin{equation} \label{A2}
 \dot f^\alpha + \hat\Omega^\alpha_{\;\;\beta} f^\beta = \psi^\alpha
\end{equation}
where $\,\hat\Omega^\alpha_{\;\;\beta}\,$ is the matrix (\ref{E21c}).

Consider now the Lorentz matrix $\, \Lambda^\mu_{\;\;\alpha}(T) = e^\mu_{(\alpha)}(T)\,$ which is a solution of equation (\ref{E21})
\begin{equation} \label{A2a} 
\dot\Lambda^\mu_{\;\;\alpha} =  \Lambda^\mu_{\;\;\beta} \hat\Omega^\beta_{\;\;\alpha} 
\end{equation}
Including that  $\,\hat\Omega_\nu^{\;\;\beta} = - \hat\Omega^\beta_{\;\;\nu}\,$, we have that the inverse matrix, $\Lambda_\mu^{\;\;\alpha} =  \eta_{\mu\nu} \Lambda^\nu_{\;\;\beta} \eta^{\beta\alpha}\,$, is a solution of 
$$ \dot \Lambda_\mu^{\;\;\alpha} + \hat\Omega^\alpha_{\;\;\beta} \,\Lambda_\mu^{\;\;\beta} = 0   $$
Hence, $\, C^\nu\,\Lambda_\nu^{\;\;\alpha}(T)\,$, with $C^\nu$ constant, is a solution of the homogeneous part of equation (\ref{A2}). The  complete equation can be solved by the method of variation of constants and we so obtain
\begin{equation}  \label{A4}
f^\alpha(T) = f^\nu_0\,\Lambda_\nu^{\;\;\alpha}(T) + \int_0^T \D t\, G^\alpha_{\;\;\nu}(T,t) \, \psi^\nu(t) 
\end{equation}
where 
\begin{equation}  \label{A3a}
 G^\mu_{\;\;\rho}(\tau,t) = \Lambda_\nu^{\;\;\mu}(\tau)\,\Lambda^\nu_{\;\;\rho}(t) = G_\rho^{\;\;\mu}(t, \tau)
\end{equation} 
acts as a kind of matrix Green function and
\begin{equation} \label{A3b}  
\partial_T G^\alpha_{\;\;\nu}(T,t) =  \hat\Omega_\lambda^{\;\;\alpha} (T) \,G^\lambda_{\;\;\nu}(T,t)  
\end{equation}
This solves the first pair of equations (\ref{A1}) provided that $\vec \psi(T)$ is known.

It is worth noticing that, except in the case of one-directional motion, the matrices $ \Lambda_\nu^{\;\;\mu}(T)\,$, $\,T\in\mathbb{R}\,$ are not in general a one-parameter subgroup of Lorentz group; however the matrices $G^\alpha_{\;\;\nu}(T,t)\,$ do have the group property:
\begin{equation}   \label{A4a}
G^\alpha_{\;\;\nu} (T,t)\,G^\nu_{\;\;\lambda} (t,t^\prime) = G^\alpha_{\;\;\lambda} (T,t^\prime)  
\end{equation}
and also
$$ G^\alpha_{\;\;\nu} (T,t) = G_\nu^{\;\;\alpha} (t,T) \qquad {\rm and}  \qquad G^\alpha_{\;\;\nu} (t,t)= \delta^\alpha_{\nu}  $$

To solve the second pair of equations we first organize the unknowns $\vec g$ and $\vec\psi$ as a skewsymmetric matrix $M_{\alpha\beta}$, with 
$$ M_{ij} = \epsilon_{ijk} g^k \,, \qquad \qquad   M_{i4} = - M_{4i} = \psi^i \,, $$
so that equations (\ref{A1}) become
\begin{equation}  \label{A5}
 \dot M_{\alpha\beta} = M_{\rho\beta}\,\hat\Omega^\rho_{\;\;\alpha} + M_{\alpha\rho}\,\hat\Omega^\rho_{\;\;\beta} - \hat{F}_{\alpha\beta}
\end{equation}

It can be easily checked that, provided that $\Lambda^\mu_{\;\alpha}$ is a solution of (\ref{E21}), 
$$ M_{\alpha\beta} = M^0_{\mu\nu} \Lambda^\mu_{\;\alpha} \Lambda^\nu_{\;\beta} \,, \qquad {\rm with}  \qquad M^0_{\mu\nu} =  M^0_{\nu\mu} \mbox{ constant} \,, $$
is a solution of the homogeneous equation and, by the method of variation of constants, we obtain that the general solution of equation (\ref{A5}) is
\begin{equation}    \label{A8}
 M_{\alpha\beta}(T) =  M_{\mu\nu}^0\,\Lambda^\mu_{\;\;\alpha}(T)\,\Lambda^\nu_{\;\;\beta}(T)  - \,\int_0^T \D t\,\hat{F}_{\mu\nu}(t) \,G_\alpha^{\;\;\mu}(T,t)\,G_\beta^{\;\;\nu}(T,t) 
\end{equation}
where $G_\alpha^{\;\;\mu}$ is obtained by raising/lowering the indices in the matrix Green function (\ref{A3a}).

Finally, as $\psi^\nu(t) = M^\nu_{\;\;4}(t)\,$, equation (\ref{A4}) leads to
\begin{equation}  \label{A9}
 f^\alpha(T) = f_0^\nu \Lambda_\nu^{\;\;\alpha}(T) +  M_{\mu\lambda}^0 \, \Lambda^{\mu\alpha}(T) \,\int_0^T \D t \, \Lambda^\lambda_{\;\;4}(t)
- \int_0^T \D t\, \int_0^t \D t^\prime \, \hat{F}_{\mu\lambda}(t^\prime)\,  G^{\alpha\mu}(T,t^\prime)\,G_4^{\;\;\lambda}(t,t^\prime)
\end{equation}
where (\ref{A4a}) has been  included.

\section*{Appendix B}
Here we outline the main traits in the derivation of the commutation relations (\ref{E24}). 

From (\ref{E22b}) we have that the only non-vanishing Lie brackets among $\hat{D}_\mu$'s are
\begin{equation}  \label{C1}
\left[\hat{D}_i, \hat{D}_4\right] = \frac1{1+\vec{X}\cdot\vec{a}(T)}\,\hat{\Omega}_i^{\;\;\sigma}(T)\,\hat{D}_\sigma   
\end{equation}
wich, combined with 
$$ \hat{D}_4 \Lambda_\nu^{\;\;\beta}(T) = \frac1{1+\vec{X}\cdot\vec{a}(T)}\, \Lambda_\nu^{\;\;\rho}(T)\hat{\Omega}_\rho^{\;\;\beta}(T) $$
readily leads to
\begin{equation}  \label{C2}
\left[\mathbf{P}_\mu, \mathbf{P}_\nu \right] = 0
\end{equation}

Now, from (\ref{E23d}) we easily obtain that 
\begin{equation}  \label{C3}
 \mathbf{P}_\mu k_\beta = \eta_{\mu\beta}
\end{equation}
which, combined with the second equation (\ref{E23a}) and (\ref{C2}), immediately yields
$$ \left[\mathbf{J}_{\alpha\beta},\mathbf{P}_\mu  \right] = 2 \,\eta_{\mu[\alpha} \mathbf{P}_{\beta]}  \,, \qquad  \qquad \left[\mathbf{J}_{\alpha\beta},\mathbf{J}_{\mu\nu}  \right] = 2\,\eta_{\mu[\alpha} \mathbf{J}_{\beta]\nu} - 2\,\eta_{\nu[\alpha} \mathbf{J}_{\beta]\mu} $$
That is, the infinitesimal generators $\mathbf{P}_\mu \,$ and $\, \mathbf{J}_{\alpha\beta}\,$ span Poincaré algebra.

To derive the commutators of these with the generators $\mathbf{D}^{\alpha\beta}_t\,$, we shall use that 
\begin{equation}  \label{C4}
D_4\chi(t,T) = \delta(T-t) \qquad {\rm and} \qquad 
\mathbf{P}_\mu \Delta^\beta(T,t,\vec{X}) = \Lambda_\mu^{\;\;\beta}(t)
\end{equation}
Moreover, form (\ref{A2a}) we have that
$$ D_4\left(\frac{\delta \Lambda_\mu^{\;\;\nu}(T)}{\delta\hat{\Omega}_{\alpha\beta}(t)}\right) = \frac{\delta \Lambda_\mu^{\;\;\rho}(T)}{\delta\hat{\Omega}_{\alpha\beta}(t)}\,\hat{\Omega}_\rho^{\;\;\nu}(T) + 2\,\delta(T-t)\,\Lambda_\mu^{\;\;[\alpha}(t)\,\eta^{\beta]\nu} $$
which can be integrated to obtain:
\begin{equation}  \label{C5}
\frac{\delta \Lambda_\mu^{\;\;\nu}(T)}{\delta\hat{\Omega}_{\alpha\beta}(t)} = 2\,\chi(t,T)\,\Lambda_\mu^{\;\;[\alpha}(t)\,G^{\beta]\nu}(t,T)
\end{equation}
We also need that
\begin{equation}  \label{C6}
\left[\hat{D}_4,\frac{\delta \qquad}{\delta\hat{\Omega}_{\alpha\beta}(t)} \right] = \frac2{1+\vec{X}\cdot\vec{a}(T)}\,\delta(T-t)\, X^j\,\delta_j^{[\beta}\eta^{\alpha]\rho}\hat{D}_\rho
\end{equation}
Then, combining equations (\ref{C4}) to (\ref{C6}) we easily arrive at
\begin{equation}  \label{C7}
\left[\mathbf{P}_\mu, \mathbf{D}^{\alpha\beta}_t \right] = 0
\end{equation}
and, including (\ref{E23a}) and the fact that
\begin{equation}  \label{C8}
\frac{\delta k_\mu(T,\vec{X})}{\delta\hat{\Omega}_{\alpha\beta}(t)} = 2\,\chi(t,T)\,\Lambda_\mu^{\;\;[\alpha}(t)\,\Delta^{\beta]}(T,t,\vec{X}) \,, 
\end{equation}
we readily obtain that $\mathbf{D}^{\alpha\beta}_t k_\mu = 0 \,$ which, combined with (\ref{E23a}) and (\ref{C7}) yields
\begin{equation}  \label{C9}
\left[\mathbf{J}_{\mu\nu}, \mathbf{D}^{\alpha\beta}_t \right] = 0
\end{equation}

Finally, to calculate the commutators between pairs of generators of the kind $\,\mathbf{D}^{\alpha\beta}_t\,$, we realise that, from (\ref{E23c}) and (\ref{C8}),
\begin{equation}  \label{C10}
\mathbf{D}^{\mu\nu}_t = \frac{\delta \qquad}{\delta\hat{\Omega}_{\mu\nu}(t)} - 
\frac{\delta k^\rho(T,\vec{X})}{\delta\hat{\Omega}_{\alpha\beta}(t)} \, \mathbf{P}_\rho
\end{equation}
and therefore
\begin{eqnarray}
\left[ \mathbf{D}^{\mu\nu}_t , \mathbf{D}^{\alpha\beta}_{t^\prime} \right] & = & \left(-\frac{\delta^2 k^\sigma(T,\vec{X})}{\delta\hat{\Omega}_{\mu\nu}(t)\,\delta\hat{\Omega}_{\alpha\beta}(t^\prime)} + \frac{\delta^2 k^\sigma(T,\vec{X})}{\delta\hat{\Omega}_{\alpha\beta}(t^\prime)\,\delta\hat{\Omega}_{\mu\nu}(t)} \right)\,\mathbf{P}_\sigma \nonumber \\[2ex]
 & & - \frac{\delta k^\sigma(T,\vec{X})}{\delta\hat{\Omega}_{\alpha\beta}(t^\prime)} \,\left[\frac{\delta \qquad}{\delta\hat{\Omega}_{\mu \nu}(t)}, \mathbf{P}_\sigma \right] + 
\frac{\delta k^\sigma(T,\vec{X})}{\delta\hat{\Omega}_{\mu\nu}(t)} \,\left[\frac{\delta \qquad}{\delta\hat{\Omega}_{\alpha\beta}(t^\prime)}, \mathbf{P}_\sigma \right] \nonumber \\[2ex]   \label{C11}
 & & + \left[\frac{\delta k^\sigma(T,\vec{X})}{\delta\hat{\Omega}_{\mu\nu}(t)} \,\mathbf{P}_\sigma,  
\frac{\delta k^\rho(T,\vec{X})}{\delta\hat{\Omega}_{\alpha\beta}(t^\prime)} \, \,\mathbf{P}_\rho \right]
\end{eqnarray}
Now, as cross partial derivatives are equal, the first term in the right hand side vanishes. Furthermore, as $\,\mathbf{P}_\sigma\,$ and $\,\mathbf{D}^{\alpha\beta}_{t^\prime} \,$ commute, we have that 
$$ \left[\frac{\delta \qquad}{\delta\hat{\Omega}_{\alpha\beta}(t^\prime)}, \mathbf{P}_\sigma \right] = - 
\mathbf{P}_\sigma\left(\frac{\delta k^\rho(T,\vec{X})}{\delta\hat{\Omega}_{\alpha\beta}(t^\prime)} \right) \,\mathbf{P}_\rho $$ 
which, substituted in (\ref{C11}) yields 
$$ \left[ \mathbf{D}^{\mu\nu}_t , \mathbf{D}^{\alpha\beta}_{t^\prime} \right] = 0 $$


\begin{thebibliography}{99}
\bibitem{Einstein1907}{Einstein A,  {\em Jahrb Rad Elektr} {\bf 4} (1907) 411}
\bibitem{Kretschmann1917}{Kretschmann E, {\em Ann Phys Lpz} {\bf 53} (1917) 575}
\bibitem{Fock}{Fock V A, {\em The Theory of Space, Time and Gravitation}, McMillan (1964) }
\bibitem{Antoci2009}{Antoci S and Liebscher D E, ``The group aspect in the physical interpretation of general relativity theory'', arXiv:gr-qc/0910.2073}
\bibitem{Ellis1995}{Ellis G F R and Matravers D R, {\em Gen Relativ Gravit} {\bf 27} (1995) 777 }
\bibitem{Zalaletdinov1996}{Zalaletdinov R, Tavakol R and Ellis G F R, {\em Gen Relativ Gravit} {\bf 28} (1996) 1251}
\bibitem{Duval1993}{Duval C,  {\em Class Quantum Grav}  {\bf 10} (1993) 2217 (arXiv:0903.1641 [math-ph])}
\bibitem{Lukierski2007}{Lukierski J, Stichel P C and Zakrzewski W J, {\em Phys Lett } {\bf B650} (2007) 203  (arXiv:hep-th/0702179)}
\bibitem{Gomis2012}{Andringa1 R, Bergshoeff E, Gomis J and  de Roo M, {\em Class Quantum Grav} {\bf 29} (2012) 235020}
\bibitem{Mashhoon}{B Mashhoon, {\em Annalen der Physik} {\bf 18} (2009) 640; C Chicone and B Mashoon, {\em Phys Rev} {\bf D74} (2006) 064019}
\bibitem{Mashhoon2}{Mashhoon B, {\em Int Journal Math Phys} {\bf D14} (2005) 171; B Mashhoon, {\em Phys Rev} {\bf D90} (2014) 124031}
\bibitem{MTW}{Misner, Thorne K and Wheeler J A, {\em Gravitation}, Freeman (1972)  }
\bibitem{Synge1}{Synge J L, {\em Relativity: the special theory}, North-Holland (1965) }
\bibitem{Walker}{Walker A G, {\em Proc Roy Soc Edinburgh} {\bf 52} (1932) 345} 
\bibitem{Bini2015}{Bini D and Mashoon B, {\em Phys Rev} {\bf D 91} (2015) 084026}
\bibitem{Marzlin1993}{Marzlin K-P, {\em Gen Relativ Gravit} {\bf 26} (1994) 619 }
\bibitem{Ni1978}{Ni W-T and Zimmermann M, {\em Phys Rev} {\bf D 17}(1978) 1473}
\bibitem{Gourgoulon}{Gourgoulhon E, {\em Special relativity in general frames}, Springer 2013 }
\bibitem{Bel93}{Bel L, ``Born's group and generalized isometries'', in {\em Relativity in General. Proceedings of the Relativity Meeting'93}, J Diaz and M Lorente eds., Editions Frontières (1994)}

\bibitem{Hicks}{Hicks N, {\em Notes on differential geometry}, Van Nostrand (1965)   }

\bibitem{Einstein1913b}{Einstein A, {\em Phys Zeitschr} {\bf 14} (1913) 1249}


\end{thebibliography}
\end{document}